\documentclass[useAMS]{mn2e}
\usepackage{times,epsfig,amsmath} 
\usepackage{graphicx}
\usepackage{epstopdf}

\newcommand{\xmm}{{\it XMM-Newton}}
\newcommand{\aap}{A\&A}
\newcommand{\apj}{ApJ}
\newcommand{\apjs}{ApJSS}
\newcommand{\aj}{AJ}
\newcommand{\mnras}{MNRAS}

\title[]
{Dark matter distribution in X-ray luminous galaxy clusters with Emergent Gravity}

\author[S. Ettori et al.]
{S. Ettori$^{1,2}$, V. Ghirardini$^{1,3}$, D. Eckert$^4$, F. Dubath$^4$, E. Pointecouteau$^{5,6}$ \\
\footnotesize 
 $^1$ INAF, Osservatorio Astronomico di Bologna, via Pietro Gobetti 93/3, 40129 Bologna, Italy \\
 $^2$ INFN, Sezione di Bologna, viale Berti Pichat 6/2, I-40127 Bologna, Italy \\
 $^3$ Dipartimento di Fisica e Astronomia Universit\`a di Bologna, via Pietro Gobetti 93/2, 40129 Bologna, Italy  \\
 $^4$ Department of Astronomy, University of Geneva, ch. d'Ecogia 16, 1290 Versoix, Switzerland \\
 $^5$ CNRS; IRAP; 9 Av. colonel Roche, BP 44346, F-31028 Toulouse cedex 4, France \\
$^6$ Universit\'e de Toulouse; UPS-OMP; IRAP; Toulouse, France
  }

\begin{document}
\maketitle 

\begin{abstract}
We present the radial distribution of the dark matter in two massive, X-ray luminous galaxy clusters, Abell~2142 and Abell~2319, and compare it with the quantity predicted
as apparent manifestation of the baryonic mass in the context of the ``Emergent Gravity'' scenario, recently suggested from Verlinde (2016).
Thanks to the observational strategy of the \xmm\ Cluster Outskirt Programme (X-COP), using the X-ray emission mapped with \xmm\ and the SZ signal in the Planck survey,
we recover the gas density, temperature and thermal pressure profiles up to $\sim R_{200}$, allowing to constrain at unprecedented level the total mass through the hydrostatic equilibrium equation.
We show that, also including systematic uncertainties related to the X-ray based mass modelling, the apparent ``dark'' matter shows 
a radial profile that has a shape different from the traditional dark matter distribution, 
with larger discrepancies (by a factor 2--3) in the inner ($r<200$ kpc) cluster's regions
and a remarkable agreement only across $R_{500}$.
\end{abstract} 
 
\begin{keywords}  
  galaxies: clusters: general -- cosmology: miscellaneous -- X-rays: galaxies: clusters.
\end{keywords}

\section{Introduction}

The distribution of the gravitating mass in galaxy clusters is one of the key ingredients to use them as astrophysical laboratories and cosmological probes (see e.g. Allen, Evrard \& Mantz 2011, Kravtsov \& Borgani 2012). In the present favourite $\Lambda CDM$ scenario, galaxy clusters are dominated by dark matter (80\% of the total mass), with a contribution in the form of hot plasma emitting in X-ray and detectable through the Sunyaev-Zeldovich (SZ, Sunyaev \& Zeldovich 1972) effect (about 15\% of the total mass, i.e. $M_{\rm DM}/M_{\rm gas} \sim 4-7$) 
and the rest in stars (few per cent; see e.g. Gonzalez et al. 2013).
Although an intriguing and plausible explanation to the observed gravitational effects induced from galaxy clusters, the still unknown nature of the dark matter invites to consider alternative scenarios.

In this paper, we present and discuss the application of one such alternative model, the ``Emergent Gravity''  theory proposed recently in Verlinde (2016), 
to the mass distribution in X-ray luminous galaxy clusters. 
The ``Emergent Gravity''  theory is a theoretical framework in which spacetime and gravity emerge together from the entanglement structure of an underlying microscopic theory.
Although a description of the cosmology is not yet available for this theory, where, in the approximation used by
Verlinde, the dark energy dominates our universe and ordinary matter only leads to small perturbations, 
the use of an effective $\Lambda CDM$ background cosmology to convert from angular to physical scales is 
still a reasonable approximation at the low redshift regime where we operate.
For the $\Lambda CDM$ model, we adopt the cosmological parameters
$H_0=70$ km s$^{-1}$ Mpc$^{-1}$ and $\Omega_{\rm m} = 1 - \Omega_{\Lambda}=0.3$.
In a similar context, the ``Emergent Gravity''  theory has already shown a good capability to reproduce the observed signal
of the galaxy-galaxy lensing profiles (Brouwer et al. 2016) and the velocity dispersion profiles of eight dwarf spheroidal satellites of the Milky Way (Diez-Tejedor et al. 2016).

\begin{figure*}
\begin{center} \hbox{
\includegraphics[width=0.48\textwidth, keepaspectratio]{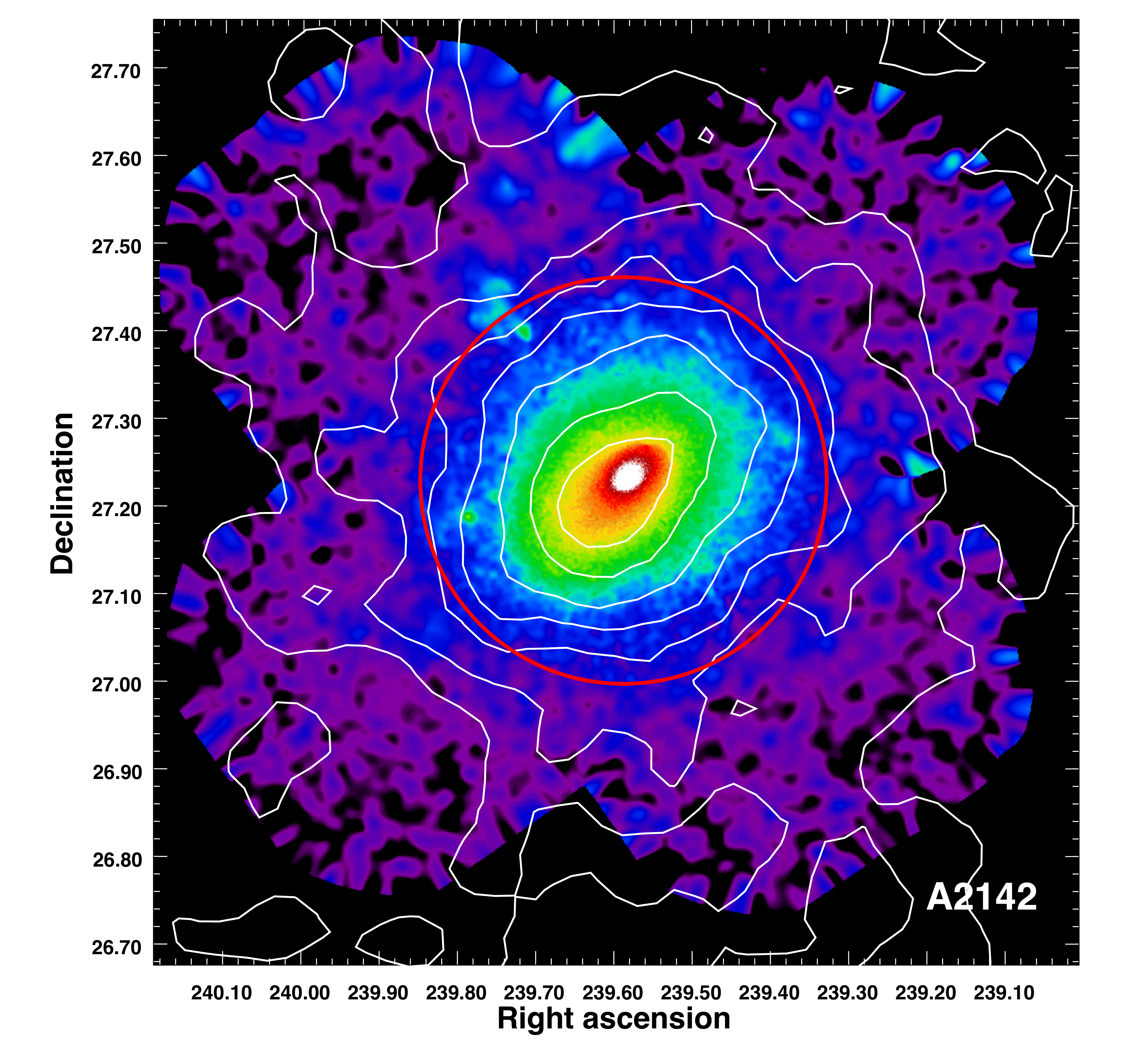}
\includegraphics[width=0.48\textwidth, keepaspectratio]{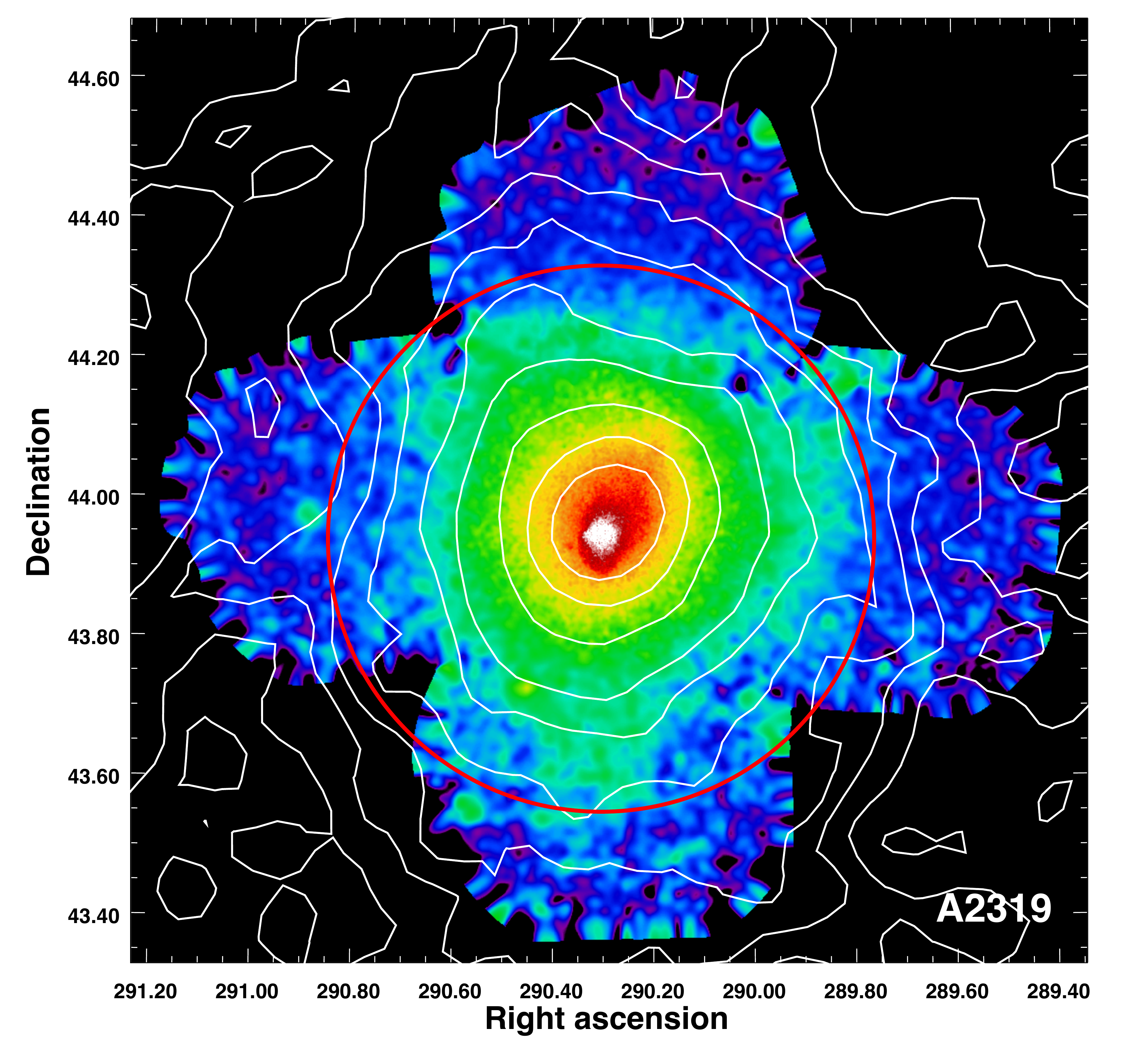}
} \end{center}
\caption{Particle background subtracted, adaptively-smoothed and vignetting-corrected \xmm\ mosaic images of X-COP clusters in the [0.7-1.2] keV band
of A2142 (left) and A2319 (right).
The corresponding Planck Compton-parameter contours are shown in white. 
The contour levels correspond to 1, 3, 5, 7, 10, 15, 20, 30, 40, and 50 $\sigma$. The red circles indicate the estimated value of $R_{500}$.
} \label{fig:obj}
\end{figure*}

In this study, we refer often to radii, $R_{\Delta}$, and masses,  $M_{\Delta}$, that are the corresponding values estimated at the given overdensity $\Delta$ 
as $M_{\Delta} = 4/3 \, \pi \, \Delta \, \rho_{\rm c,z} R_{\Delta}^3$, where $\rho_{\rm c,z} = 3 H_z^2 / (8 \pi G)$ is the critical density of the universe 
at the observed redshift $z$ of the cluster, and $H_z = H_0 \, \left[\Omega_{\Lambda} +\Omega_{\rm m}(1+z)^3\right]^{0.5}$ is the value of the Hubble constant at the same redshift.

The paper is organized as follows. In Section~2, we describe the ``Emergent Gravity'' scenario and how an apparent dark matter distribution 
can be associated to the observed baryonic mass. 
In Section~3, we present the dark matter profiles reconstructed through techniques based on X-ray and SZ data only 
in two massive galaxy clusters that are part of the X-COP sample, an \xmm\ Large Program which targets the outer regions of a sample of 13 massive clusters in the redshift range $0.04-0.1$ at uniform depth. 
In Section~4, we compare these dark matter profiles with the ones recovered though ``Emergent Gravity'', assessing the systematic uncertainties affecting the X-ray mass measurements, 
and summarize our main findings in Section~5.
Unless mentioned otherwise, the quoted errors are statistical uncertainties at $1 \sigma$ confidence level.

\section{Apparent dark matter in the Emergent Gravity}

In the `Emergent Gravity'', dark matter can appear as manifestation of an additional gravitational force describing the ``elastic'' response due to the entropy displacement, and with a strength that can be described in terms of the Hubble constant and of the baryonic mass distribution for a spherically symmetric, static and isolated astronomical system as 
(equation~7.40 in Verlinde 2016): 
\begin{equation}
\int_0^r \frac{G\, M_{\rm DM, EG}^2(r')}{r'^2}dr' = \frac{M_{\rm B}(r) \, cH_0 \, r}{6}.
\label{eq:verlinde}
\end{equation}
By operating the derivatives with respect to the radius of the two terms, and rearranging the quantities to isolate the dark matter component $M_{\rm DM}$, 
it is straightforward to show that the following relation holds:
\begin{eqnarray}
M_{\rm DM, EG}^2(r) & = & \frac{cH_0}{6G} r^2 \frac{d(M_{\rm B}(r) \,r)}{dr} \nonumber \\
  & = & \frac{cH_0}{6G} r^2 \left(M_{\rm B}(r) +r \frac{dM_{\rm B}(r)}{dr} \right) \nonumber \\
  & = & \frac{cH_0}{6G} r^2 \left(M_{\rm B}(r) +4\pi r^3 \rho_{\rm B}(r)\right) \nonumber \\
  & = & \frac{cH_0}{6G} r^2 M_{\rm B}(r) \left(1 + 3 \delta_{\rm B} \right),
\label{eq:mdm_eg}
\end{eqnarray}
where $M_{\rm B}(r) = \int_0^r 4\pi \rho_{\rm B} {r'^2}dr' = M_{\rm gas}(r) +M_{\rm star}(r)$ is the baryonic mass equal to the sum of the gas and stellar masses,
and $\delta_{\rm B}$ is equal to $\rho_{\rm B}(r) / \bar{\rho}_{\rm B}$, with $\bar{\rho}_{\rm B} = M_{\rm B}(r)/V(<r)$ representing the mean baryon density within the spherical volume $V(<r)$.
In our case, the gas mass has been obtained from the integral over the cluster's volume of the gas density that is obtained from the geometrical deprojection 
of the observed surface brightness (Fig.~\ref{fig:obj}) including a careful treatment of the background subtraction. This allows to resolve the signal out to about $R_{200}$.
The stellar mass has been estimated using a Navarro-Frenk-White (NFW, Navarro et al. 1997) profile with a concentration of 2.9 (see e.g. Lin et al. 2004) 
and by requiring the $M_{\rm star}(<R_{500}) / M_{\rm gas}(<R_{500}) = 0.39 \, \left(M_{500}/10^{14} M_{\odot}\right)^{-0.84}$ (Gonzalez et al. 2013).  

It is worth noticing that Eq.~\ref{eq:mdm_eg} can be expressed as an acceleration $g_{\rm EG}$ depending on the acceleration $g_{\rm B}$ induced from the baryonic mass
\begin{eqnarray}
g_{\rm EG} & = & G \frac{M_{\rm DM, EG} +M_{\rm B}}{r^2}  \nonumber \\
  & = & g_{\rm B} \, \left( 1 + y^{-1/2} \right),
\label{eq:mond}
\end{eqnarray}
where $y = 6 / (c H_0) \times g_{\rm B} / (1 + 3 \delta_{\rm B})$. Equation~\ref{eq:mond} takes a form very similar to the one implemented in MOND (e.g. Milgrom \& Sanders 2016) with a characteristic acceleration $a_0 = c H_0 (1 + 3 \delta_{\rm B}) / 6$.

\begin{figure*}
\begin{center} \hbox{
\includegraphics[width=0.48\textwidth, keepaspectratio]{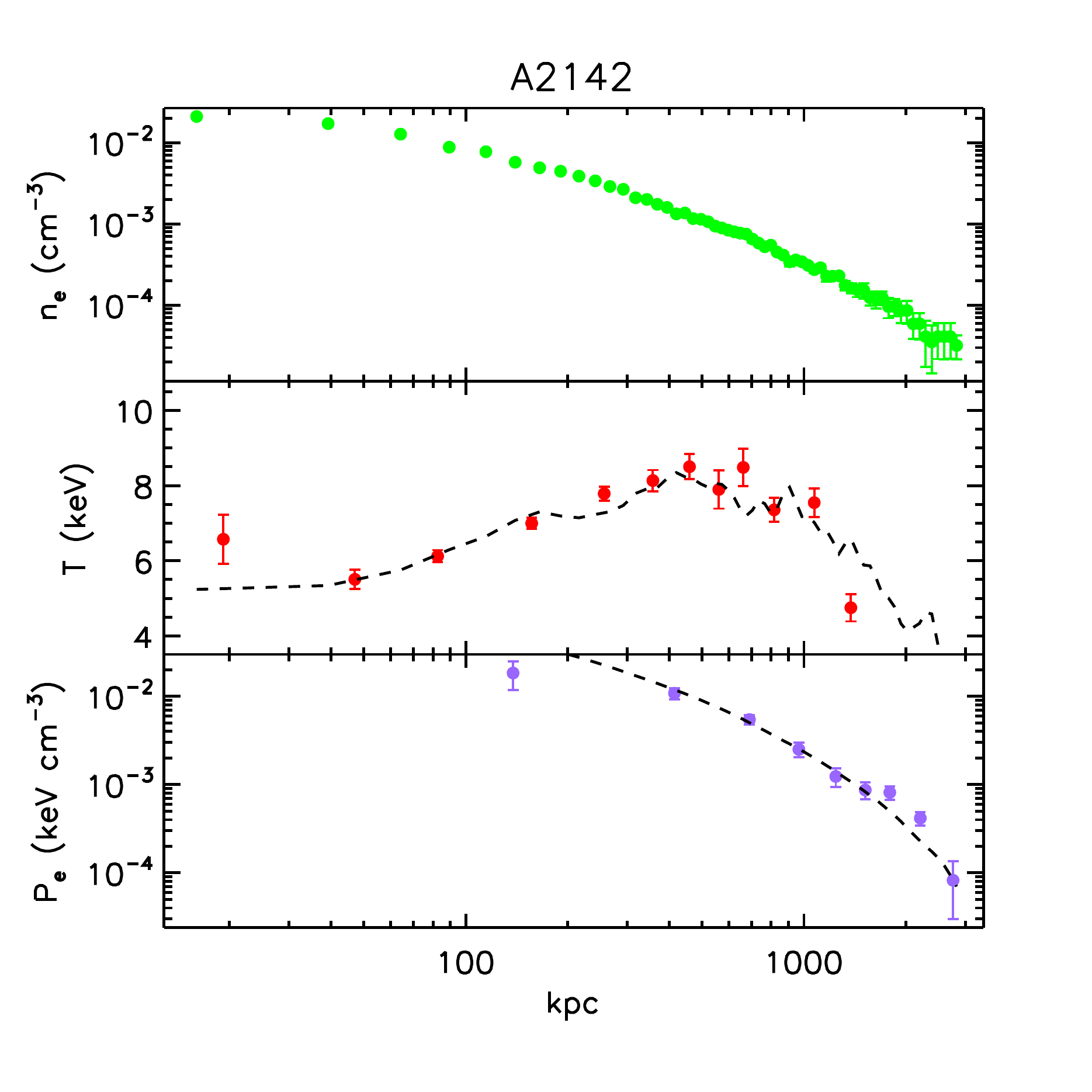}
\includegraphics[width=0.48\textwidth, keepaspectratio]{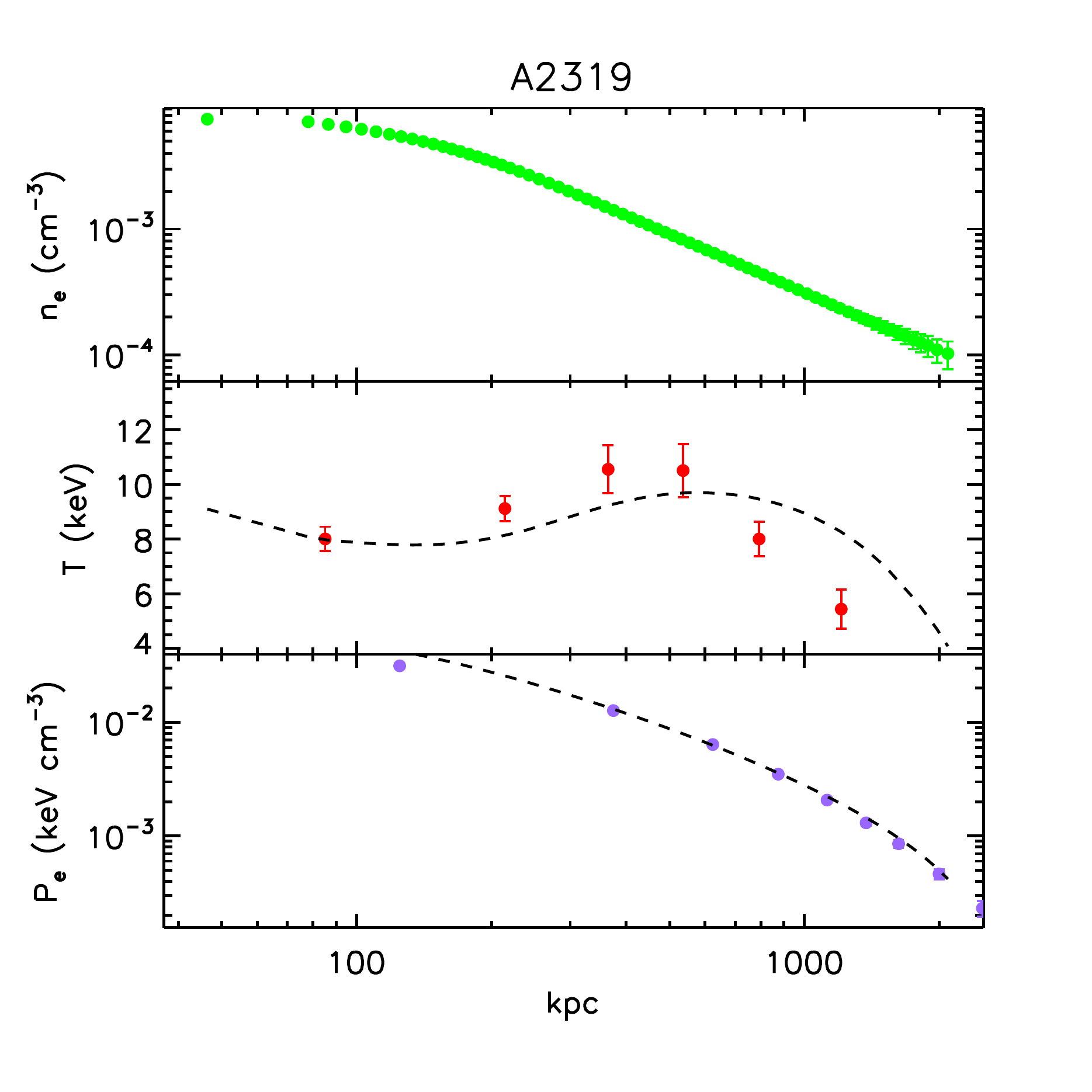}
} \end{center} \vspace*{-1cm}
\caption{(From top to bottom) Observed deprojected electron density, temperature and SZ pressure profiles, with the statistical error bars overplotted.
The dashed lines indicate the temperature and pressure profiles required from the best-fitting NFW mass model for the given gas density profile (see Sect.~3).
% with the best-fit models used to reconstruct the hydrostatic mass with the {\it forward} method.
} \label{fig:tn}
\end{figure*}

\section{Dark matter with the Hydrostatic Equilibrium Equation}

We evaluate how the apparent dark matter profile described in eq.~\ref{eq:mdm_eg} reproduces 
the mass distribution recovered by using the hydrostatic equilibrium equation applied
to two massive, X-ray luminous galaxy clusters that are part of the X-COP sample. 
The \xmm\ Cluster Outskirts Project (X-COP; Eckert et al. 2016) has been built to target the outer regions of a sample of 13 massive clusters 
($M_{500} > 3\times10^{14} M_\odot$) in the redshift range $0.04-0.1$ at uniform depth. 
The sample was selected based on the signal-to-noise ratio in the Planck SZ survey (Planck Collaboration et al. 2011) 
with the aim of combining high-quality X-ray and SZ constraints throughout the entire cluster volume. 
Our observing strategy allows us to reach a sensitivity of $3\times10^{-16}$ ergs cm$^{-2}$ s$^{-1}$ arcmin$^{-2}$ in the [0.5-2.0] keV range 
thanks to a good control of systematic uncertainties. 
The two objects in exam, Abell~2142 and Abell~2319, are the first targets of the X-COP sample for which the complete \xmm\ analysis 
of their gas properties out to $R_{200}$ has been completed (see Fig.~\ref{fig:obj}).
Abell~2142 ($z=0.091$) shows a relatively relaxed morphology extended along the SE/NW axis, and is undergoing some minor mergers in its outskirts (Owers et al. 2011; Eckert et al. 2014).
This cluster was mapped in the framework of X-COP pilot project (Tchernin et al. 2016).
Abell~2319 ($z=0.056$, Struble \& Rood 1999) is also a massive system in which the galaxy distribution indicates 
that it is a merger of two main components with a 3:1 mass ratio, the smaller system being located $\sim10'$ north of the 
main structure (Oegerle et al. 1995). The cluster exhibits a prominent cold front SE of the main core (Ghizzardi et al. 2010) 
and a giant radio halo (Farnsworth et al. 2013; Storm et al. 2015). 
This is one of the most significant SZ detections in the Planck catalogue (Planck Collaboration et al. 2014) 
and its complete X-ray analysis, combined with the SZ pressure profile and resolved in 8 azimuthal sectors,  will be presented in a forthcoming paper (Ghirardini et al. in prep.). 
Considering the merging state of this galaxy cluster, we present here the analysis performed in the most relaxed sector, the one enclosed between Position Angles 180$^\circ$ and 225$^\circ$. 
Under a reasonable approximation, these clusters are following Verlinde's prescriptions for the validity of the EG modelling: they are reasonably spherical, quite isolated 
(being not embedded in the potential well of any neighbour objects and with no major mass accretion), and with the largest baryonic component, 
the hot plasma mapped in X-ray and SZ bands, in hydrostatic equilibrium.

The physical quantities directly observable are the density $n_{\rm gas}$ and temperature $T_{\rm gas}$ of the X-ray emitting gas, and the SZ pressure profile $P_{\rm gas}$.
The gas density is obtained from the geometrical deprojection of the X-ray surface brightness in Fig.~\ref{fig:obj}.
Thanks to the observational strategy implemented in X-COP, we are able to correct the X-ray emission for the presence of clumps 
both by masking substructures spatially resolved with \xmm\ and by measuring the azimuthal median, instead of the azimuthal mean, out to $\sim 1.2 R_{200}$,
with a median relative uncertainty of 6\% and 1\% in A2142 and A2319, respectively.
The estimates of the gas temperature are based on the modelling with an absorbed thermal component of the \xmm\ spectra extracted from 
concentric annuli around the X-ray peak in the [0.5--12] keV energy band and corrected from the local sky background components (see Tchernin et al. 2016 for details).
A typical statistical error lower than 5\% is associated to these spectral measurements, with a profile resolved in 12 bins out to 1.4 Mpc in A2142 and in 14 bins out to 1.9 Mpc in A2319.
The SZ electron pressure profile is obtained from the deprojection of the azimuthally-averaged integrated Comptonization parameter $y$ 
extracted from a re-analysis of the SZ signal mapped with Planck (e.g. Tchernin et al. 2016, Planck Collaboration et al. 2013) 
and that extends up to $\sim$ 3 and 4 Mpc in A2142 and A2319, respectively.
The electron density, temperature and SZ pressure profiles are presented in Fig.~\ref{fig:tn}.

\begin{figure*}
\hbox{
\includegraphics[width=0.47\textwidth, keepaspectratio]{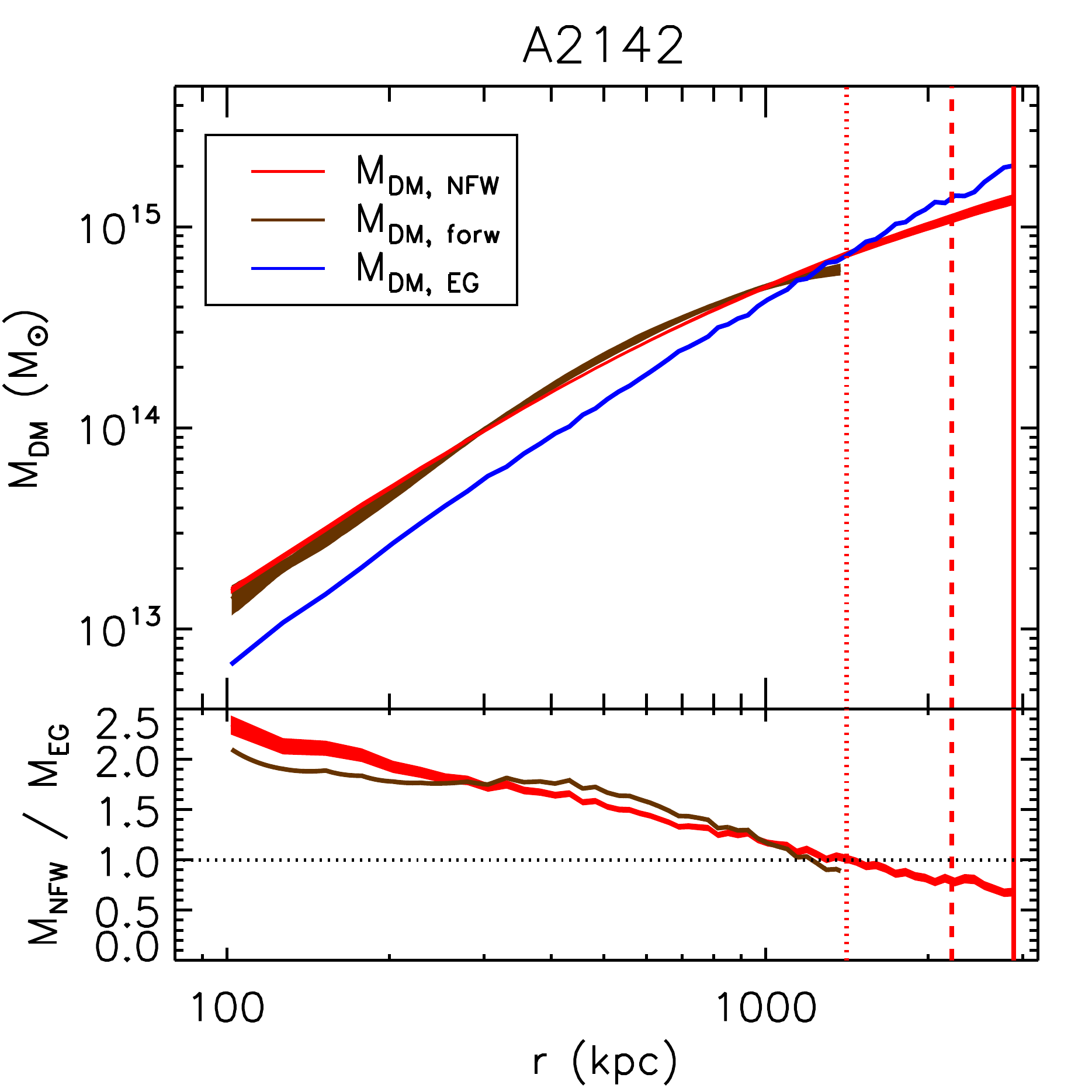}
\includegraphics[width=0.47\textwidth, keepaspectratio]{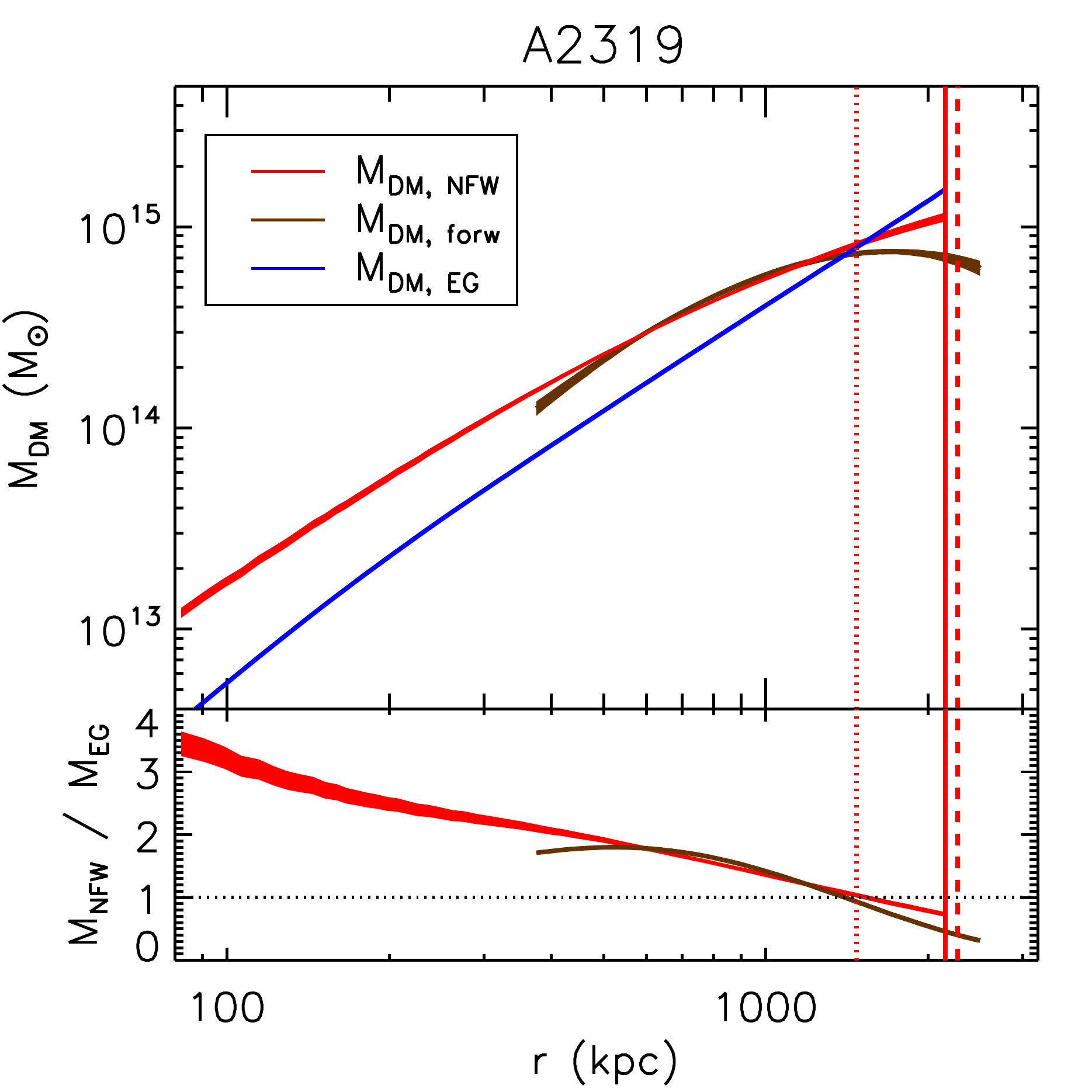}
} \caption{Dark matter profiles obtained using (i) the {\it backward} method with a NFW mass model; 
(ii) the {\it forward} method by fitting with functional forms the gas density profile and either the deprojected temperature profile (A2142) or the SZ pressure profile (A2319).
In the latter case, the mass profiles are shown only within the radial range where the data are fitted.
The dark matter profiles (blue curve) predicted from the ``Emergent Gravity'' framework as obtained from equation~\ref{eq:mdm_eg} are also shown. 
% (blue curve: $M_{\rm B} = M_{\rm gas} +M_{\rm star}$; purple curve: $M_{\rm B} = M_{\rm gas}$ ).
The thickness of the lines shows the statistical uncertainty associated to the best-fit mass model.
Dotted/dashed/solid lines indicate $R_{500}$/$R_{200}$/outermost radius of the extracted gas density profile, respectively, as estimated in the X-ray analysis.
In the bottom panel, the ratio between the NFW mass model and $M_{\rm DM, EG}$ is shown.
} \label{fig:mass}
\end{figure*}

Under the assumption that the intracluster medium has a spherically-symmetric distribution and follows the perfect gas law 
($P_{\rm gas} = k T_{\rm gas} n_{\rm gas}$, where $k$ is the Boltzmann's constant,  and $n_{\rm gas}$ is the sum of the electron and proton densities 
$n_{\rm e} + n_{\rm p} \approx 1.83 n_{\rm e}$),
the gas density, combined with the X-ray spectral measurements of the gas temperature and/or the SZ derived gas pressure,
allows to evaluate the total mass within a radius $r$ through the hydrostatic equilibrium equation (see e.g. Ettori et al. 2013)
\begin{equation}
M_{\rm tot}(<r) = - \frac{r \, P_{\rm gas}}{\mu m_{\rm u} G \, n_{\rm gas}} \frac{d \log P_{\rm gas}}{d \log r}, 
%   + \frac{\partial \log n_{\rm gas}}{\partial \log r} \right),
   \label{eq:mhe}
\end{equation}
where $G$ is the gravitational constant, $m_{\rm u} = 1.66 \times 10^{-24}$ g is the atomic mass unit, 
and $\mu=0.61$ is  the mean molecular weight in atomic mass unit.
In this analysis, we have applied both the {\it backward} and the {\it forward} method. In the {\it backward} method, a parametric mass model is assumed 
and combined with the gas density profile to predict a gas temperature profile that is then compared, through e.g. a $\chi^2$ minimization, with the one either measured 
in the spectral analysis or estimated as SZ $P_{\rm gas}/n_{\rm gas}$ (losing the spatial resolution in the inner regions because of the modest 7 arcmin FWHM angular 
resolution of our Planck SZ maps, but gaining in radial extension due to the Planck spatial coverage; Planck Collaboration et al. 2013) to constrain the mass model parameters. 
Here, we combine both sets of constraints by summing up  
$\chi^2_T = \sum_i^{N_x} \left( T_i - T_{\rm mod, i} \right)^2 / \epsilon_{T, i}^2$, 
that is estimated from the spectral measurements of the gas temperature (and relative errors $ \epsilon_T$) resolved in $N_x$ radial bins, 
and $\chi^2_{SZ} = \Delta^T C^{-1} \Delta$, 
that is evaluated from the SZ pressure profile resolved in $N_{SZ}$ radial bins, by defining the elements of the matrix $\Delta$ as $\Delta_{ij} = P_i - P_{\rm mod, j}$ and properly weighting by its covariance matrix $C$.
% for A2142, where we have proven that the combination of the different observed profiles provides 
% consistent results on the estimates of the hydrostatic masses (Tchernin et al. 2016). 
% In the case of A2319, considering the systematics that might affect the spectral measurements above 1 Mpc (Ghirardini et al. in prep.), and 
% the large signal-to-noise ratio in SZ available well beyond $R_{200}$, we combine the gas density and the SZ pressure.
In the present analysis, we adopt a NFW mass model with two free parameters, the mass concentration and $R_{200}$.
This mass model provides a better representation (i.e. lower $\chi^2$) of our data than any mass model including a central core.
The statistical error associated to the mass is evaluated at each radius considering the range of the mass values allowed from the distribution of the 
best-fitted parameters within a $\Delta \chi^2 = 2.3$. The temperature and pressure profiles required from the best-fitting NFW mass model are shown in Fig.~\ref{fig:tn}.
In the {\it forward} method, some functional forms are fitted to the gas density profile and the deprojected gas temperature (or pressure) profile.
The hydrostatic equilibrium equation (eq.~\ref{eq:mhe}) is then directly applied to evaluate the radial distribution of the mass. 
The errors are estimated through a Monte-Carlo process.
The functional forms used to reproduce the profiles are a double $\beta-$model for the gas density (Cavaliere \& Fusco-Femiano 1976), 
a 6-parameters function for the temperature, $T = p_0 \, (p_3 +(r/p_1)^{p_4}) / (1 +(r/p_1)^{p_4}) /( 1+(r/p_2)^2)^{p_5}$ (e.g. Vikhlinin et al. 2006, Baldi et al. 2012), 
or 5-parameters generalized NFW for the pressure, $P = p_0 / \left( (r/p_1)^{p_2} (1+ (r/p_1)^{p_3}\right)^{(p_4-p_2)/p_3}$ (e.g. Arnaud et al. 2010).

\section{Results on the dark matter mass profiles}

From equation~\ref{eq:mhe}, using a {\it backward} method with a NFW model, we measure in A2142 a total mass of
$M_{500} = 8.7 \times 10^{14} M_{\odot}$, with a relative statistical error of 3 per cent, and
$R_{200} = 2211 \pm 47$ kpc, with the gas density that extends up to $r =2890$ kpc.
As discussed in Tchernin et al. (2016), the hydrostatic mass profile agrees well with the one obtained by weak lensing and caustics measurements out to $R_{200}$.
In A2319, we measure $M_{500} = 7.5 \times 10^{14} M_{\odot}$, with a relative statistical error of 2 per cent, and $R_{200} = 2084 \pm 13$ kpc, 
with the outermost radius for the gas density at 3 Mpc.
A systematic uncertainty of about 10 per cent on these mass measurements is estimated by applying the {\it forward} method (with both the temperature and pressure profiles). 
The dark matter distribution is then $M_{\rm DM} = M_{\rm tot} - M_{\rm B}$, where $M_{\rm B}$ is the baryonic mass estimated as described in Section~2.

In Figure~\ref{fig:mass}, we show the mass profiles obtained both in a context of a $\Lambda CDM$ model and following the prescriptions for an emergent dark matter contribution.
An encouraging match between the two mass profiles is obtained at $r \approx R_{500}$, where we measure $M_{\rm DM}/M_{\rm DM, EG} = 1.01 \pm 0.04$
in A2142 and $0.81 \pm 0.02$ in A2319, where the errors include only the propagation of the statistical uncertainties. 
On the contrary, $M_{\rm DM, EG}$ underpredicts significantly, by up to a factor of $2-3$, the requested amount of matter 
to maintain the hydrostatic equilibrium in the central regions, $r<200$ kpc.
We conclude that, although the total masses within $\approx R_{500}$ are in good agreement, the overall shape of the DM profiles looks quite different, 
with EG lacking some NFW-type curvature.

By inverting the hydrostatic equilibrium equation, and assuming as boundary condition $P_{\rm out} = P(R_{500})$, 
we can also estimate the gas temperature profiles that the computed $M_{\rm DM, EG}$ would imply for the measured gas density profiles. 
% We show the comparison of this temperature profile with the ones both measured in the spectroscopic analysis and required from the best-fitting mass model in Fig.~\ref{fig:tt}. 
The tension below 1000 kpc can then be translated in a difference in the gas temperature of 2--4 keV, that can be hardly accommodated with the present observational constraints \footnote{By comparing the predicted and the observed temperature profiles, we estimate a $\Delta \chi^2$ between $\sim$200 (for A2319) and 830 (A2142) in disfavour of the $M_{\rm DM, EG}$. The null hypothesis that the NFW model, with two free parameters, does not provide a better representation of the data than EG, with no free parameters, is excluded at $>99$ per cent).}.

\section{Conclusions}

We have investigated the dark matter profiles in two massive X-ray luminous galaxy clusters for which the gas density and temperature (from \xmm\ X-ray data) and
SZ pressure profiles (from Planck) are recovered at very high accuracy up to about $R_{200}$. 
By applying the hydrostatic equilibrium equation on these profiles, we constrain the dark matter distribution using different methods 
and models, obtaining results consistent within $\sim 10$ per cent.
Other systematic uncertainties might affect our mass reconstruction, such as any other (e.g. non-thermal) contribution to the total gas pressure (e.g. Nelson et al. 2014b, Sereno et al. 2017),
other terms that account for departures from the hydrostatic equilibrium (e.g. Nelson et al. 2014a, Biffi et al. 2016), or the violation of the assumed sphericity of the gas distribution (e.g. Sereno et al. 2017).
All these contributions have been shown to affect more significantly the clusters' outskirts and tend to bias higher (by 10-30\%) the total mass estimates at $r>R_{500}$, with lower effects in the inner regions. 
However, in A2142, we observe an excellent agreement between the reconstructed mass profiles using X-ray, weak lensing and galaxy dynamics (Tchernin et al. 2016), suggesting that, at least for this system,
the hydrostatic equilibrium is a valid approximation allowing a robust constraint of the mass profile out to $R_{200}$.

Then, we compare those to $M_{\rm DM, EG}$, the value predicted to play the role of an apparent dark matter as manifestation of an excess of gravity in the ``Emergent Gravity'' scenario suggested in Verlinde (2016), that has the appealing property to depend only on the observed baryonic mass and the Hubble constant, with no extra free parameter. 
To this aim, we recover the baryonic mass as the sum of the observed gas mass and of the statistically estimated mass in stars. 
We observe that $M_{\rm DM, EG}$ reproduces well the dark matter distribution requested to maintain the gas in pressure equilibrium beyond 1 Mpc 
from the cluster core, with a remarkable good match at $r \approx R_{500}$, but presents significant discrepancies (by a factor $2-3$) in the innermost 200 kpc. 

We note that any underestimate of the hydrostatic mass (in the order of 10 per cent or less, if any, in the latest analyses of samples of galaxy clusters 
-e.g. Mahdavi et al. 2013, Donahue et al. 2014, Applegate et al. 2016, Smith et al. 2016; bias that we exclude in A2142 as discussed in Tchernin et al. 2016)
would imply a higher true mass at larger radii shifting the radius at which $M_{\rm DM, EG}$ and the expected dark matter value agree.
Considering the extremely tight constraints on the gas density that come from the exquisite combination of high statistics and control of the systematics in the background modelling,
the only way to reconcile this discrepancy would require a systematic over-estimate of the gas temperature by 2--4 keV at $r<1000$ kpc, 
that is completely inconsistent with the present observational constraints, also accounting for potential systematics due to the calibration of the X-ray instruments 
(e.g. Schellenberger et al. 2015).  
Otherwise, this discrepancy might suggest that some temperature (or gas entropy) contribution, with an effect comparable with a modulation by some scale radius 
and larger in the inner cluster's regions, is still missing in the Verlinde's formula. Massive (probably sterile) neutrinos can also accomodate this tension (e.g. Nieuwenhuizen 2016).

A larger sample of high-quality data, as the ones that will be available in the X-COP project in the next future, 
will improve the statistical constraints on the reliability of any alternative scenario, as the ``Emergent Gravity'' 
here discussed, to the dark matter.

\section*{ACKNOWLEDGEMENTS} 
We thank the anonymous referee for helpful comments that improved the presentation of the work.  
This research has received funding from the European Union's Horizon 2020 Programme under AHEAD project (grant agreement n. 654215).
We thank Guillaume Hurier to have generated the Planck SZ maps used in X-COP.
SE acknowledges the financial support from contracts ASI-INAF I/009/10/0, 
% PRIN-INAF 2012 ``A unique dataset to address the most compelling open questions about X-Ray Galaxy Clusters'', 
NARO15 ASI-INAF I/037/12/0 and ASI 2015-046-R.0.

\end{document}